\documentclass[12pt]{article}
\usepackage[english]{babel}
\usepackage[T1]{fontenc}
\usepackage{amssymb}    
\usepackage{latexsym}
\usepackage{amsfonts}
\usepackage{epsfig}
\usepackage{dsfont}

\begin{document}
\def\op#1{\mathcal{#1}}
\def\bfnull{\relax{\rm O \kern-.635em 0}}
\def\dop{{\rm d}\hskip -1pt}
\def\mez{\frac{1}{2}}
\def\sx{\left}
\def\dx{\right}
\def\na{\nabla}
\def\imez{\frac{i}{2}}
\def\a{\alpha}
\def\b{\beta}
\def\g{\gamma}
\def\d{\delta}
\def\e{\epsilon}
\def\ve{\varepsilon}
\def\t{\theta}
\def\l{\lambda}
\def\m{\mu}
\def\n{\nu}
\def\pg{\pi}
\def\r{\rho}
\def\s{\sigma}
\def\t{\tau}
\def\z{\zeta}
\def\c{\chi}
\def\p{\psi}
\def\o{\omega}
\def\G{\Gamma}
\def\D{\Delta}
\def\T{\Theta}
\def\L{\Lambda}
\def\Pg{\Pi}
\def\S{\Sigma}
\def\O{\Omega}
\def\pb{\bar{\psi}}
\def\cb{\bar{\chi}}
\def\lb{\bar{\lambda}}
\def\i{\imath}
\def\Pii{\mathcal{P}}
\def\Q{\mathcal{Q}}
\def\K{\mathcal{K}}
\def\A{\mathcal{A}}
\def\N{\mathcal{N}}
\def\F{\mathcal{F}}
\def\Gi{\mathcal{G}}
\def\Ci{\mathcal{C}}
\def\oL{\overline{L}}
\def\oM{\overline{M}}
\def\wk{\widetilde{K}}
\def\hb{\overline{h}}
\def\eq#1{(\ref{#1})}
\newcommand{\be}{\begin{equation}}
\newcommand{\ee}{\end{equation}}
\newcommand{\ba}{\begin{eqnarray}}
\newcommand{\ea}{\end{eqnarray}}
\newcommand{\ban}{\begin{eqnarray*}}
\newcommand{\ean}{\end{eqnarray*}}
\newcommand{\nn}{\nonumber}
\newcommand{\nin}{\noindent}
\newcommand{\fgl}{\mathfrak{gl}}
\newcommand{\fu}{\mathfrak{u}}
\newcommand{\fsl}{\mathfrak{sl}}
\newcommand{\fsp}{\mathfrak{sp}}
\newcommand{\fusp}{\mathfrak{usp}}
\newcommand{\fsu}{\mathfrak{su}}
\newcommand{\fp}{\mathfrak{p}}
\newcommand{\fso}{\mathfrak{so}}
\newcommand{\fg}{\mathfrak{g}}
\newcommand{\fr}{\mathfrak{r}}
\newcommand{\fe}{\mathfrak{e}}
\newcommand{\rE}{\mathrm{E}}
\newcommand{\rSp}{\mathrm{Sp}}
\newcommand{\rSO}{\mathrm{SO}}
\newcommand{\rSL}{\mathrm{SL}}
\newcommand{\rSU}{\mathrm{SU}}
\newcommand{\rUSp}{\mathrm{USp}}
\newcommand{\rU}{\mathrm{U}}
\newcommand{\rF}{\mathrm{F}}
\newcommand{\R}{\mathbb{R}}
\newcommand{\C}{\mathbb{C}}
\newcommand{\Z}{\mathbb{Z}}
\newcommand{\Hb}{\mathbb{H}}
\def\oL{\overline{L}}
\def\mW{\mathcal{W}}

%
\def\Jnote#1{{\bf[JL: #1]}}
\def\Snote#1{{\bf[SV: #1]}}


\begin{titlepage}
\begin{center}

\hfill hep-th/0605063

\rightline{\small ZMP-HH/2006-06}
\rightline{\small DESY 06-058}
\vskip 1cm
                    
{\Large \bf $N=1$ domain wall solutions of massive type II
supergravity as generalized geometries}

\vskip 1 cm

{\bf Jan Louis$^{a,b}$ and
Silvia Vaul\`a$^{a,c}$ }
\
\vskip 0.8cm

{}$^{a}${\em II. Institut f{\"u}r Theoretische Physik der Universit{\"a}t Hamburg\\
Luruper Chaussee 149,  D-22761 Hamburg, Germany}\\
 {\tt jan.louis@desy.de} 
\vskip 0.4cm

{}$^{b}${\em Zentrum f\"ur Mathematische Physik, 
Universit\"at Hamburg,\\
Bundesstrasse 55, D-20146 Hamburg}

\vskip 0.4cm

{}$^{c}${\em DESY \\
Notkestra\ss e,  D-22761 Hamburg, Germany}\\
 {\tt silvia.vaula@desy.de} 
\vskip 0.4cm

\end{center}

\vskip 1cm

\begin{center} {\bf ABSTRACT } \end{center}

\noindent
We study $N=1$ domain wall solutions of type IIB supergravity
compactified on a Calabi--Yau manifold in the presence of RR and NS
electric and magnetic fluxes. We show that the dynamics of the scalar
fields along the direction transverse to the domain wall is described by
gradient flow equations controlled by a superpotential $W$. We then
provide a geometrical interpretation of the gradient flow equations in terms
of the mirror symmetric compactification of type IIA. 
They correspond to a set of 
generalized Hitchin flow equations of a manifold with 
${\rm SU}(3) \times {\rm SU}(3)$ structure which is fibered  over 
the direction transverse to the domain wall.

\vfill

\bigskip

\noindent May 2006

\end{titlepage}

        

\section{Introduction} 

Domain wall (DW) solutions of supergravity
have received
a lot of attention recently which is largely due to their role in
the AdS/CFT  correspondence \cite{Aharony:1999ti}. However, apart from
this application they also have  been studied as a class of
supersymmetric ground
states alternative to the commonly considered  Minkowski or AdS backgrounds.
In particular supergravities with non-trivial background fluxes often do
not admit a stable, four-dimensional supersymmetric ground state
but they do have BPS DW solutions.
For example, in type IIB supergravity compactified on Calabi-Yau
threefolds with non-trivial three-form fluxes 
it is necessary to include D-branes and
orientifold planes in order to cancel the tadpoles induced by the
fluxes and to obtain an $N=1$ supersymmetric
Minkowski background \cite{Giddings:2001yu}.
On the other hand without orientifold planes 
no four-dimensional  Minkowski background is allowed.
However, in this case  three-dimensional  
$N=1$ DW solutions do exist \cite{Behrndt:2001mx,Mayer:2004sd}.

In this paper we continue the study of such DW solutions of type IIB and
generalize the previous results \cite{Behrndt:2001mx,Mayer:2004sd}  
in various respects.
More specifically we start from type  IIB supergravity compactified on
Calabi-Yau threefolds with electric and
magnetic background
three-form flux for both the NS three-form $H_3$ and the RR three-form $F_3$
\cite{Michelson}--\cite{Louis:2002ny}.
In the presence of the magnetic fluxes the four dimensional
antisymmetric tensors fields $B_{\mu\nu}$ and $C_{\mu\nu}$ become
massive \cite{Louis:2002ny}.
For this case the corresponding supergravity has only recently been
constructed in  refs.\ \cite{Dall'Agata:2003yr}--\cite{Sommovigo:2005fk}. 
Using these four-dimensional $N=2$
supergravities we study their $N=1$ DW solutions including non-trivial magnetic fluxes. 
We find that the resulting DW necessarily is
flat and furthermore that the background profile of the scalar fields is governed
by a set of gradient flow equations expressed in terms of a single
superpotential $W$, which is related
to the superpotential suggested in~\cite{Taylor:1999ii}.

The DW solutions of type IIB  have their mirror analogous in type IIA.
Without fluxes mirror symmetry identifies type IIB  compactifications
on a Calabi-Yau manifold $\tilde Y$ with type IIA compactified on the
mirror Calabi-Yau $Y$ \cite{HKT}. In the presence of  
RR fluxes mirror symmetry is straightforwardly extended by also
exchanging the respective flux parameters \cite{Taylor:1999ii,Louis:2002ny}.
For NS fluxes the situation is slightly more involved in that mirror
symmetry can relate
Calabi-Yau compactification with fluxes to  purely geometrical
compactification on a manifold $\hat Y$ without flux 
\cite{Vafa:2000wi}--\cite{FMT} 
or possible also to non-geometrical
backgrounds \cite{nongeo}. For the case of geometrical backgrounds 
$\hat Y$ is no longer a Calabi-Yau manifold but rather a manifold 
with $SU(3)$ structure or more generally with $SU(3)\times
SU(3)$-structure
\cite{HitchinHF}--\cite{Benmachiche:2006df}.
In such compactifications the
(intrinsic) torsion of $\hat Y$ plays the `mirror-role' of the fluxes.

This generalized mirror symmetry is also reflected in the DW
solutions. For electric fluxes it was shown in  \cite{Mayer:2004sd}
that the mirror symmetric DW can be interpreted as a solution of
type IIA supergravity in a warped background $M^{1,2}\times_w X_7$.
As a consequence of the $N=1$ supersymmetry of the DW $X_7$
has $G_2$ holonomy and furthermore consists of a six-dimensional 
manifold $\hat Y$ fibered over
the real line. The $G_2$ holonomy constrains $\hat Y$ to be within a
 special class of manifolds with
$SU(3)$ structure termed `half-flat' \cite{HitchinHF,CS}.
{}From a mathematically point of view such fibration were studied
in \cite{HitchinHF} and the DW solution precisely corresponds
to the Hitchin flow equations.

For magnetic fluxes mirror symmetry is more involved.
In \cite{Grana:2005ny,GLW} it is shown that in this case $\hat Y$
has to be within a special class of manifolds  
with $SU(3)\times SU(3)$ structure.
In this paper we generalize the analysis of ref.~\cite{Mayer:2004sd}
and show that the mirror symmetric DW solution of type IIB with
magnetic NS-flux also is of the form $M^{1,2}\times_w X_7$.
However, in this case  $\hat Y$ has to be 
a manifold with $SU(3)\times SU(3)$ structure which satisfies
a set of generalized Hitchin flow equations given in ref.\
\cite{JW}. 
$X_7$ in turn has an integrable $G_2\times G_2$ structure and is 
Ricci-flat as demanded by string theory.

This paper is organized as follows. In section~2 we set the stage for
our analysis and recall the $N=2$ supergravity arising as the low
energy limit of type IIB string theory compactified on Calabi-Yau
threefolds with background flux. In section~3.1 we study the $N=1$
DW solutions and show that the scalar fields vary according to  
gradient flow equations. In section~3.2
we explicitly solve these equations and rewrite the solution in terms of
mirror symmetric type IIA variables. This sets the stage for section~4
where we show that the DW solutions correspond to generalized Hitchin
flow equations of a geometrical $SU(3)\times SU(3)$ background.
Further details are found in two appendices.

\section{$N=2$ Supergravity with Abelian electric and magnetic
charges}\setcounter{equation}{0}\label{sugrapar}
In order to set the stage for the discussion of the DW solutions let
us briefly recall the structure of $N=2$ supergravity with massive
tensor multiplets as it arises from type IIB string theory
compactified on Calabi-Yau threefolds with both electric and magnetic 
three-form fluxes.
The $N=2$ supergravity including massive tensor multiplets has been
constructed in references
\cite{Dall'Agata:2003yr,Sommovigo:2004vj,D'Auria:2004yi,Sommovigo:2005fk}
where further details can be found.
Here (and in appendix \ref{th}) we only summarize the results needed in the
following.

An $N=2$  tensor multiplet contains $n_T\leq 3$ antisymmetric tensor, $4-n_T$
real scalars and 
two Weyl fermions as its components. If the tensors are massless   they can
be dualized into scalars and hence a massless tensor multiplet is dual
to a hypermultiplet which contains four real scalars and two Weyl
fermions \cite{deWit:1982na,Theis:2003jj}. In this dual formulation the remnant of the tensors are
translational isometries acting on the dual scalars. In
the standard (ungauged) $N=2$ supergravity \cite{Andrianopoli:1996cm}
one dualizes all tensor multiplet such that the theory contains only
one gravitational multiplet, vector multiplets and 
hypermultiplets.

On the other hand a massive tensor is dual to a massive
vector and 
it is often more convenient to keep the tensor multiplet in the
spectrum. Such a theory can be viewed as a
$N=2$ supergravity with tensor multiplets which is 
deformed by Abelian electric and magnetic charges
\cite{Dall'Agata:2003yr,Sommovigo:2004vj,D'Auria:2004yi}. These
charges are not related to any gauging of isometries on the residual
scalar manifold. Instead the electric charges appear in Green--Schwarz
type interaction of the tensors with the gauge fields while the
magnetic charge appear in the St\"uckelberg mass terms of the tensors.

In this paper we do not discuss the general case \cite{D'Auria:2004yi}
but instead focus on type IIB theories
compactified on Calabi-Yau threefolds $\tilde Y$ 
in the presence of electric and
magnetic three-form fluxes
\cite{Sommovigo:2004vj,Dall'Agata:2001zh,Louis:2002ny}. In this case the
spectrum features a gravitational multiplet
\be(g_{\m\n},\,\p_{A\m},\,\p^A_\m,\,A^0_\m)\ee
where $g_{\m\n}$ is
the metric, $\p_{A\m},\, A=1,2$ are the two chiral gravitinos while $A^0_\m$
is the graviphoton. In addition there are $n_V=h^{(1,2)}$ vector multiplets 
\be(A^i_\m,\,\l^{iA},\,\l^i_A,\,t^i)\ ,\qquad i = 1,\ldots, n_V\ ,\ee 
where 
$A^i_\m$ are  the gauge bosons, $\l^{iA}$ are the doublets of chiral gaugini
while $t^i$ are complex scalars.\footnote{Here $h^{(1,1)}$ and $h^{(1,2)}$
are the Hodge numbers of the Calabi-Yau manifold $\tilde
Y$. Throughout the paper we denote by $t^i$ the scalars of the vector
multiplets irrespective of their geometric origin. In type IIB
compactification they correspond to deformations of the complex
structure while in type IIA compactification they  parameterize 
the K\"ahler deformation (cf. appendix \ref{th}).}
Finally there are $n_H=h^{(1,1)}$ hypermultiplets and one double tensor
multiplet. Since they couple non-trivially it is convenient to combine
them as
\ba (\zeta_\alpha,\,\zeta^\alpha,\,q^u,\,B_{I\m\n})\ , && \quad\alpha =
1,\ldots,2n_H+2\ , \\
&&\quad  u = 1,\ldots, 4n_H+2,\quad I=1,\,2\ .\nn \ea
Each of these multiplets features two chiral hyperinos which
we collectively denote as $\zeta_\alpha$.
The bosonic components of the hypermultiplets are $4n_H$ real scalars,
while the double tensor multiplet contains two antisymmetric tensors
$B_{I\m\n}$ (they are the four-dimensional part of the RR and the NSNS
two--forms) together with the axion $l$  and the four dimensional
dilaton $\varphi$. We denote the scalars in the hypermultiplets and
in the double tensor multiplet collectively by $q^{u}$. 

The background fluxes arise from expanding both the RR three-form
$F_3$ and the NS three-form $H_3$ along the third cohomology $H^3$ 
of the Calabi-Yau manifold
\be
F_3 + \tau H_3 = m^\Lambda \alpha_\Lambda - e_\Lambda \beta^\Lambda\ ,\qquad
\Lambda = 0, \ldots, h^{(1,2)}\ ,
\ee
where
\be \label{fluxes}
e_\L=e_\L^1+\t e_\L^2 \ , \qquad m^\L=m^{\L1}+\t m^{\L2}\ ,
\ee
are the electric and magnetic background fluxes\footnote{The
nomenclature electric--magnetic is linked to the definition of the
electric versus magnetic gauge bosons which arise in the expansion of
the type IIB four form $C_4$ according to
$C_4 = A_1^\Lambda \alpha_\Lambda - \tilde A_{1\,\Lambda}
\beta^\Lambda +\ldots$. Here $A_1^\Lambda$ are  the $(h^{(1,2)}+1)$
electric gauge bosons (including the graviphoton) while $\tilde
A_{1\,\Lambda}$ are the corresponding dual magnetic gauge bosons.}
and $\t$ is the ten-dimensional complex type IIB dilaton $\t=l+ie^{-\phi}$.
The three-forms $(\alpha_\Lambda, \beta^\Lambda)$ denote a real,
symplectic basis of $H^3$. 

In the next section we will search for $N=1$ DW solutions of the effective
supergravity arising from type IIB compactifications. 
For this task we need the scalar part of the
supersymmetry transformation of the fermions which can be non--trivial
along the DW. In particular, in the following we set to zero the field strengths of the
vectors and the tensors.
With this assumption the supersymmetry transformation of the two
gravitinos $\p_{\m A}$ has the form  \cite{Dall'Agata:2003yr,Sommovigo:2004vj,D'Auria:2004yi}
\be
\d\p_{\m A}=D_\m\ve_A+iS_{AB}\g_\m\ve^B\label{psi} \ , 
\ee
where $\ve_A$ are the two supersymmetry parameters and $S_{AB}$ is a
hermitian $SU(2)$ matrix which depends on the background fluxes
\eq{fluxes}.
For the type IIB compactifications under consideration one finds \cite{Sommovigo:2004vj}
\be\label{shiftS}
S_{AB} =\frac{ i}{2}\, \sigma^x_{AB}\, \o^x_I\ \langle V,\K^I\rangle \
, \quad I=1,2\ ,
\ee
where the quaternionic connection $\o^x_I$ is given by \cite{Ferrara:ik}
\be\label{ofs}
\o_1^x= \delta^{x3}e^{2\varphi}\ , \qquad
\o_2^{1}=- e^{2\varphi} Im\t\ ,\quad \o_2^{2}=0\ ,
\quad\o_2^{3}=e^{2\varphi} Re\t\ .
\ee
Here $e^{2\varphi} = \frac1{8}e^{-K_H} e^{2\phi} $  is the
four-dimensional real dilaton, $\phi$ is the ten-dimensional IIB dilaton
and $\frac1{8} e^{- K_H}$ is the volume of $\tilde Y$ which is defined in \eq{KH3}.
We also 
assembled the background fluxes  
into (symplectic) vectors $\K^I=(m^{I\L},e^I_\L)$ and defined the 
symplectic inner product $\langle\,,\,\rangle$ as: 
\be\label{Vdef}
\langle V,\K^I\rangle =(L^\Lambda e_\Lambda^I - M_\Lambda m^{\Lambda I})\ ,
\ee
where 
$V=(L^\L, M_\L)$ is defined in (\ref{Vdef2}).
The electric and magnetic charges are not arbitrary, as supersymmetry
in four dimensions \cite{Sommovigo:2004vj} and the tadpole
cancellation condition in ten dimensions \cite{Taylor:1999ii}, impose 
\be\langle\K^1,\K^2\rangle=0\ .\label{tadpole}\ee
Inserting (\ref{ofs}) into (\ref{shiftS}) $S_{AB}$ reads explicitly
\be\label{shift}
S_{AB}=\frac{i}{2}\,e^{2\varphi}\Big[\s^{3}_{AB}\,\big(\langle V,\K^1\rangle
+ \langle V,\K^2\rangle
\textrm{Re}\t\big)-\langle V,\K^2\rangle\, \textrm{Im}\t\,\s^{1}_{AB}\Big]\ .
\ee

The supersymmetry transformations of the gaugini
are given by
\be
\d\l^{iA}=i\partial_\m t^i\g^\m\ve^A+W^{iAB}\ve_B\label{la}\ ,
\ee
where
\ba\label{shiftW}
W^{iAB} & = & 
 ig^{i\bar \jmath}\sigma^{AB}_x \o^x_I\langle U_{\overline{\jmath}},\K^I\rangle
\\
&=&i g^{i\bar\jmath}e^{2\varphi}\left[\s_{3}^{AB}(\langle
U_{\bar\jmath},\K^1\rangle+ \langle U_{\bar\jmath},\K^2\rangle
\textrm{Re}\t)-\langle U_{\bar\jmath},\K^2\rangle
\textrm{Im}\t\,\s_{1}^{AB}\right]\ ,\nn
\ea
and we defined $U_i=\nabla_iV \equiv
(\partial_i  + \frac12 \partial_iK_V)V$ where $K_V$ is the K\"ahler
potential of the vector multiplets defined in \eq{KV}.

Finally the supersymmetry transformations of the hyperinos read
\be
\d\z_\a=iP_{u A\a}\partial_\m q^u\g^\m\ve^A+N_\a^A\ve_A\label{zi}\ ,
\ee
where 
\be
N^{\alpha}_A  = 
-2\mathcal{U}^\alpha_{AI}\langle V,\K^I\rangle\ .
\label{shiftN}
\ee
The matrixes $P_{u A\a}$ play the r\^ole of a vielbein on the scalar
manifold spanned by the $q^u$'s, while $\mathcal{U}^\alpha_{AI}$ are
remnants of the vielbein on the quaternionic manifold along the
directions which have been dualized into scalars (see
appendix~\ref{th} and in particular reference \cite{Dall'Agata:2003yr}
for more details).  



\section{$N=1$ Domain Wall
solutions}\label{DW}\setcounter{equation}{0}

\subsection{Gradient Flow Equations}
After this brief review of the $N=2$ low energy supergravity arising in
Calabi-Yau compactifications of type IIB string theory let us now turn
to
the main topic of this paper and study its three-dimensional $N=1$ DW
solutions. That is we study solutions of the four-dimensional $N=2$ supergravity
which preserve the three-dimensional Lorentz group $SO(1,2)$ and half
of the supercharges. We split the  coordinates $x^\mu, \mu = 0,\ldots,
3$ of the four-dimensional space-time into coordinates $(x^m,z),
m=0,\,1,\,2,$
where $x^m$ denote the coordinates along the DW while $z$
parameterizes the direction normal to the DW. Accordingly we split the
background metric preserving Lorentz invariance as 
\be 
g_{\m\n}(x^\m)\,dx^\mu dx^\nu = 
e^{U(z)}\hat{g}_{mn}(x^m)\,  dx^mdx^n+g_{zz}(z)\,dzdz \ .
\ee
where $\hat{g}_{mn}(x^m)$ is the metric of a three-dimensional
space-time which we assume to have constant curvature. (In the
following the `hatted' quantities will refer to the three-dimensional 
un-warped metric.)
Furthermore, following \cite{Behrndt:2001mx,Mayer:2004sd} we choose to
parameterize $g_{zz}(z)= - e^{-2pU(z)}$
where $p$ is an arbitrary real number. Finally using $\mu=e^{U(z)}$ instead of
$z$ as the coordinate of the transverse space we arrive at 
\be\label{edef}
g_{\m\n}(x^\m)dx^\mu dx^\nu = \m^2 \hat{g}_{mn}(x^m)dx^mdx^n-\frac{d\m d\m}{\m^2
{\mW}^2(z)}=\eta_{\a\b}e^\a_{\,\m}e^\b_{\,\n}dx^\mu dx^\nu \ ,
\ee
where \be\label{defmu}\mW=\pm e^{pU(z)} U^\prime(z)\ .\ee
The non--vanishing components of the vierbein defined in \eq{edef} take the form
\be \label{eDW}
e^a_m=\hat{e}^a_m e^{U(z)}\ ,\qquad e^3_z=e^{-pU(z)}\d^3_z\ ,
\quad a=0,\,1,\,2\ ,
\ee
while the non-vanishing components of the spin connection $\omega$ are
found to be 
\be\label{oDW}
\o^{ab}_m=\hat\o^{ab}_m\ ,\qquad\o^{a3}_m=e^{(p+1)U(z)}\,U^\prime(z)
\hat{e}^a_m\ .
\ee 

Since we are interested in DW solutions which preserve four
supercharges we first study the supersymmetry transformations of the
fermionic fields. More precisely we solve 
$\delta_{\epsilon}\textrm{fermions} = 0$ for half of the
supercharges. This is most easily done by imposing from the very beginning a relation  
on the two supersymmetry parameters $\varepsilon_A, A=1,2$ which reads
\cite{Behrndt:2001mx,Mayer:2004sd}
\be\ve_A=\hb A_{AB}\g_3\ve^B\label{mezzospin}\ .\ee
Here $h(z)$ is a complex function while $A_{AB}$ is a constant matrix.
Consistency of \eq{mezzospin} with its hermitian conjugate implies
$h\hb=1$ while $A_A^{\ B}\equiv A_{AC}\,\e^{CB}$ must be a hermitian
matrix which in addition satisfies
\be A_A^{\ B}A_B^{\ C}=\d_A^C\ .\ee
Thus $A$ has to be a suitable linear combination of $(\mathds{1},
\vec\s)$
where $\vec\s$ are the Pauli matrices.

Finally, the condition of constant curvature of $\hat g_{mn}(x^m)$ 
can also be expressed as the integrability condition of 
\cite{Behrndt:2001mx,Coussaert:1993jp}
\be\label{CC}
\hat{D}_m(h^\frac12\ve_A)=\frac i\ell \hat e^a_m\g_a\g_3h^\frac12\ve_A\ee
where $\frac{1}{\ell^2}$ is the three dimensional cosmological constant.

The next step is to look for solutions of 
\be\d\p_{\m A}=\d\l^{iA}=\d\z_\a=0\label{0}\ee
with \eq{mezzospin} imposed. Furthermore, we only allow the scalar
fields to be non-trivial in the DW background setting all other fields
to zero. Since we are most interested in the values of the scalars
transverse to the DW 
we suppose in the following that they  only depend on the 
coordinate $z$ and ignore their $x^m$ dependence.

Let us  first consider $\d\p_{Am}=0$. 
Using \eq{eDW}--\eq{CC} one derives
\be
A^{AB}D_m\ve_B=-\left(\frac{i}{\ell}e^{-U}+\frac12
e^{(p+1)U}U^\prime\right)\hb\g_m\ve^A\ . \ee
Inserted into $\d\p_{Am}=0$ one obtains
\be
\left(\frac{i}{\ell}e^{-U}+\frac12
e^{(p+1)U}U^\prime\right)\hb\g_m\ve^A = i A^{AB} S_{BC} \gamma_m
\ve^C\ .
\label{pippo}\ee
This implies that $A^{AB} S_{BC}$
 is proportional to the identity or in other words
\be  \label{Wdef}
i A^{AB} S_{BC} = \frac12 W\delta^A_C \ ,\ee
where the proportionality factor defines the superpotential $W$.
{}From \eq{shift} and \eq{Wdef} we infer the structure of $ A_A^{\ B}$
to be
\be \label{Aprop}
 A_A^{\ B}=\frac{1}{\sqrt2}(-\s^{1\ B}_A+\s^{3\ B}_A)\ ,\ee
and the constraint
\be\label{condtau1}(Im\t-Re\t)\langle V,\K^2\rangle=\langle
V,\K^1\rangle \ .\ee
Inserting \eq{Wdef} - \eq{condtau1} into \eq{pippo} we finally arrive at 
\ba
\frac i\ell\!\!\!&=&\!\!\! \frac14 e^{U}( hW-\bar h\bar W)\ ,\label{cosm}\\
U^\prime\!\!\!&=&\!\!\!\frac12 e^{-pU}( hW+\bar h\bar W)\ ,\label{cosm1}\\
W\!\!\!&=&\!\!\! 4e^{\varphi}e^{K_H/2}\langle
V,\K^2\rangle\ .\label{Wresult}\ea
We see that the cosmological constant is determined by the imaginary
part of $hW$ while the derivative of the warp factor is determined by
the real part. $W$ itself is determined by the fluxes.

Before continuing let us briefly discuss
the limiting cases of only RR fluxes ($\K^2=0$) and only NS fluxes
($\K^1=0$). In the first case we see from \eq{shift} and \eq{Wdef} that $A$ is
proportional to $\sigma^3$ and no consistency condition needs to be
imposed. For only NS fluxes the consistency condition is 
\be Im\t-Re\t=0\ .\ee

Now we look at to the solution of $\d\p_{Az}=0$. 
It turns out that we can follow precisely the same steps as done in
\cite{Behrndt:2001mx} with the only difference that we have to use 
the constraint \eq{condtau1}. Suppressing the intermediate steps we
arrive at
\be
\label{cosm2}\hb D_zh=\frac{2i}{\ell}\,e^{-(p+1)U}\ .
\ee

The solution of $\d\l^{iA}=0$ proceeds analogously. We insert
\eq{mezzospin} and \eq{shiftW} into \eq{la} and obtain the constraint 
\be\label{condtau2}(Im\t-Re\t)\langle U_i,\K^2\rangle=\langle
U_i,\K^1\rangle \ .\ee 
Differentiating \eq{condtau1} with respect to $z$ and
using \eq{condtau2} we conclude 
\be Im\t-Re\t=\a\label{fixtau}\ ,\ee
where $\a$ is a real constant ($\a=0$ holds if and only if there are no RR fluxes).
For our purposes we do not need to find an explicit solution for the
constraints \eq{condtau1}, \eq{condtau2}. Note that they are
satisfied, for instance, if  the two vectors of electric/magnetic
charges are parallel $\K^1=\a\K^2$ where $\a$ is defined in equation
\eq{fixtau}. This is consistent with the tadpole cancellation
condition \eq{tadpole} and with the two limiting cases $\K^1=0$ or $\K^2=0$.

Inserting \eq{condtau2} into \eq{la} we obtain the flow equations for
the scalars $t^i$
\be\label{condz}\partial_zt^i=-
g^{i\bar\jmath}e^{-pU(z)}\hb\,\nabla_{\bar\jmath} \overline W\ .\ee

The analysis of $\d\z_\a=0$ proceeds analogously and one inserts
\eq{shiftN} and  \eq{mezzospin} into \eq{zi}. Using the quaternionic relations \eq{quatrel1}--\eq{quatrel4} one finds
\be\label{condq}
\partial_zq^u=- g^{uv}e^{-pU(z)}\hb\,\partial_v\overline W\ ,
\ee
where $g^{uv}$ is defined in \eq{quatm}. 
In addition one finds that $hW$ has to be real
 or in other words $\bar h$ is determined as the phase of $W$.
\be \label{zerocosm}
\bar{h}=\frac{W}{|W|}\ .\ee
This in turn implies that the cosmological constant on the DW must be
zero, as we can see from equation \eq{cosm}. Therefore  the metric $\hat{g}_{mn}$ on the DW is flat:
\be ds^2=\m^2\eta_{mn}dx^mdx^n-\frac{d\m d\m}{\m^2\mathcal{W}^2}\ .
\label{metrsol}\ee
Note that \eq{metrsol} holds for $\K_1=0$ and in particular also for
$\K_2=0$, that is in the case where just RR fluxes are
present~\cite{Mayer:2004sd}. 

Using \eq{zerocosm} 
we can insert 
\eq{cosm1} into \eq{defmu} to arrive at
\be \mW(z)=\pm hW=\pm|W|\ .\label{dabliu}\ee

Using as a transverse coordinate $\m(z)=e^{U(z)}$, \eq{condz} and \eq{condq} can be written as gradient flow equations 
\ba
\m\frac{dt^i}{d\m}&=&-
g^{i\overline\jmath}\nabla_{\overline\jmath}\ln\overline W\label{attz}\ ,\\
\m\frac{dq^u}{d\m}&=&- g^{uv}\partial_v\ln\overline W\ .\label{attq}
\ea



\subsection{Solutions of the flow equations}

So far we derived the gradient flow equations for  an $N=1$ BPS domain
wall  in type IIB
supergravity compactified on a Calabi-Yau manifold $\tilde Y$ 
in the presence of electric and
magnetic RR and NS fluxes. The purpose of this section is to study
their solutions and to prepare for a geometrical interpretation in a
mirror symmetric compactification of type IIA on some generalized
manifold $\hat Y$.  


We will not consider the most generic solution but instead
follow \cite{Mayer:2004sd} and restrict the space of scalar
fields which can vary along the DW. More precisely the scalars in the
vector multiplets $t^i$ and the four-dimensional dilaton $\varphi$ 
can be non-trivial along the DW. As we discuss in appendix~\ref{th}, 
half of the scalars in the hypermultiplets are geometrical moduli of
the Calabi-Yau manifold. In type IIB compactifications they correspond
to deformations of the K\"ahler form and we denote them by 
$z^a= \s^a + i\lambda^a$. Following \cite{Mayer:2004sd} we only 
allow the $\lambda^a$ to be non-trivial in the DW solution while $\s^a$
together with the remaining scalar fields from the RR sector are kept constant.

Let us first focus on the flow equations for the hypermultiplet
scalars. Inserting \eq{Wresult} and \eq{zerocosm} into
\eq{cosm1} and \eq{condq} we arrive at
\ba
&&\partial_z q^u=-2e^{-pU+\varphi+\frac{K_H}{2}}g^{uv}\partial_v (2\varphi+K_H)|\langle
V,\K^2\rangle|\label{dequ}\ ,\\
&&\partial_z U(z)=4e^{-pU+\varphi+\frac{ K_H}{2}} |\langle
V,\K^2\rangle|\ ,\label{dezu}
\ea
where $K_H$ is defined in \eq{KH2}  and \eq{KH3}.
Comparing \eq{dezu} and \eq{dequ} one obtains
\be\frac{dq^u}{dU}=-\frac12 g^{uv}\partial_v(2\varphi+K_H)\label{quatatt}\ . \ee
This equation shows that  the $U$--dependence of the quaternionic
fields is not modified by the magnetic fluxes and thus we expect that the
solution coincides with the solution derived in \cite{Mayer:2004sd}.

In order to solve equation \eq{quatatt} let us first note that on the
submanifold spanned by the scalars $\varphi$ and
$\lambda^a$ the inverse metric $g^{uv}$ is block diagonal with the
components
\be
g^{\varphi\varphi} = 1 \ , \qquad
g^{a{b}}=-\frac23\left(d\,d^{ab}-3\l^a\l^b\right)\ ,\label{met}
\ee 
where we have evaluated $g^{ab}$ in the large volume limit
and defined 
\be
d=d_{abc}\l^a\l^b\l^c\ , \qquad
d_a=d_{abc}\l^b\l^c\ ,\qquad
d_{ab}=d_{abc}\l^c \ ,
\ee
with $d^{ab}$ being the inverse of $d_{ab}$.
Inserting \eq{met}  into \eq{quatatt} we obtain the solution
\be e^{\varphi}=C\,e^{-U(z)}\ , \qquad
z^a=i\l^a=iD^a\,e^{2U(z)}\label{kmod}\ ,\ee
where $C$ and $D^a$ are integration constants.
{}From \eq{KH3} we learn 
\be
 e^{- K_H}= \frac43De^{6U(z)}\ ,\label{sol1}
\ee
where we abbreviated $ D=d_{abc}D^aD^bD^c$.
Note that as expected \eq{kmod} and \eq{sol1} coincide with the result of reference \cite{Mayer:2004sd}.


Let us now consider the vector multiplets scalars. Also in this case 
it is more convenient to consider \eq{condz} instead of  \eq{attz}
which, following \cite{Behrndt:2001mx}, we rewrite as follows
\be\partial_z\begin{pmatrix}{Y^\L-\overline{Y}^\L\cr \mathcal{F}_\L-\overline\mathcal{F}_\L }\end{pmatrix}=-i4 e^{(1-p)U+\varphi+\frac{ K_H}{2}}\begin{pmatrix}{m^{\L}\cr e_\L}\end{pmatrix}\label{hitch}\ee
where we have suppressed the label ``2'' on the NSNS fluxes and defined
\be \label{newsect}{\mathcal V}\equiv h\,e^{U(z)}V=
h\,e^{U(z)}\begin{pmatrix}{L^\L\cr
M_\L}\end{pmatrix}\equiv\begin{pmatrix}{Y^\L\cr
\mathcal{F}_\L}\end{pmatrix} \ .\ee
Using the solution \eq{sol1}, choosing $D=12C^2$ and performing the change of coordinates defined by 
\be\label{cochange0} e^{(p+3)U(z)}\partial_z=\partial_w\ee
equation \eq{hitch} becomes
\be\partial_w\begin{pmatrix}{Y^\L-\overline{Y}^\L\cr
\mathcal{F}_\L-\overline\mathcal{F}_\L
}\end{pmatrix}=-i\begin{pmatrix}{m^\L\cr
e_\L}\end{pmatrix}\ .\label{hitch1}\ee
If we set $p=-3$ and $m^{\L}=0$ we recover the result of
\cite{Mayer:2004sd}.

In order to derive further useful relations, let us display
\eq{hitch1} more explicitly.
Using \eq{specialc} and the normalization $Y^0=-\frac
i2$ we infer 
\be b^i=-2\textrm{Im}Y^i\ , \qquad v^i= 2\textrm{Re}Y^i\
,\label{x2}\ee
where we split $\L = 0,i$. Inserted into \eq{hitch1} using
\eq{ppotvect} we arrive at
\ba
&& 0=m^{0}\\
&& \partial_w b^i= m^{i}\label{hb}\\
&&\frac12c_{ijk}\partial_w(v^jv^k)-\frac12c_{ijk}\partial_w(b^jb^k)=e_i\label{hi}\\
&&-\frac12c_{ijk}\partial_w(b^iv^jv^k)+\frac16c_{ijk}\partial_w(b^ib^jb^k)=e_0\label{h0}
\ea
Solutions of equations \eq{hb}--\eq{h0} are discussed in appendix \ref{soluz}.

Note that equations
\eq{cosm} and \eq{cosm1} can be rewritten in terms of the rescaled section $\mathcal{V}$
\be \langle {\rm Re}\mathcal{V},\K_2\rangle= e^{(p+2)U} \partial_wU\ , \qquad
\langle {\rm Im}\mathcal{V},\K_2\rangle = 0\ .
\label{dezuz}
\ee
Using \eq{x2}, \eq{tadpoleK}, \eq{m0} and \eq{k0} one can easily check the second equation in \eq{dezuz}
and compute  the first to be
\be e^{(p+2)U} \partial_wU
=\frac12(v^ie^2_i+c_{ijk}v^ib^jm^{2k})\label{partU}\ .\ee
Multiplying \eq{hi} by $v^i$ and making use of \eq{hb} one can derive by comparison with \eq{partU} 
\be 
e^{-K_V}\equiv\frac43\,c_{ijk}v^iv^jv^k=4e^{2U}\ ,\label{volume}\ee
where we also used \eq{KV2}.
Note that the final form of $K_V$ does not depend
on the presence of the magnetic fluxes and therefore coincides
with the results  of \cite{Mayer:2004sd}. 
Let us also observe at this point that the ten-dimensional type IIA dilaton $\phi_A$
defined by $ e^{2\phi_A} =\frac18 e^{2\varphi-K_V}$ is given by the
integration constant introduced in \eq{kmod} 
$e^{\phi_A}=C$, as can be seen 
from \eq{kmod} and \eq{volume}. This will be important in the next section.

We are now in the position to formulate the DW gradient flow equations in
a very compact way, in terms of the quantities $(Z^A,\,W_A)$ and
$(X^\L,\,F_\L)$ introduced in appendix \ref{th}.
First notice that the relation between 
$(X^\L,\,F_\L)$ and the sections $(Y^\L,\,\mathcal{F}_\L)$ can be
deduced from equations \eq{hsc}, \eq{newsect} and \eq{volume}. In particular, setting the irrelevant overall phase to zero, that is $h=1$, we obtain
\be\begin{pmatrix}{X^\L\cr
F_\L}\end{pmatrix}=2\begin{pmatrix}{Y^\L\cr\mathcal{F}_\L}\end{pmatrix}\
,\ee 
and as a consequence \eq{hitch1} now reads
\be\partial_w\begin{pmatrix}{{\rm Im}X^\L\cr {\rm Im }
F_\L}\end{pmatrix}=-\begin{pmatrix}{m^\L\cr e_\L}\end{pmatrix}\
.\label{dRephi+}\ee
Furthermore, in these variables \eq{dezuz} reads
\be{\rm Im}X^\L e_\L- {\rm Im}F_\L m^\L=0\ .\label{dImphi+}\ee

Let us return to the flow equations for the hypermultiplet scalars
\eq{attq} or \eq{dequ} respectively, whose solution we already gave in
\eq{kmod}. However, in order to compare the solution with the Hitchin
flow equation of the next section it is useful to rewrite them in a
form similar to \eq{dRephi+}. This is achieved in terms of rescaled  
variables $(Z^A,\, W_A)_\eta$ given by
\be\label{Zrescale} (Z^A,\,W_A)=|c|\, (Z^A,\,W_A)_\eta\ , \qquad |c|^2\equiv
e^{K_V-K_H}=\frac D3 e^{4U}\ ,\ee
where the last equality used \eq{sol1} and \eq{volume}.
The geometrical meaning of this rescaling will become more transparent
in the next section.

Recalling the definition \eq{scH}, the solution \eq{kmod} and the
gradient flow equation \eq{dezuz}, one can easily check that 
\be\partial_w\begin{pmatrix}{{\rm Im}Z^A\cr {\rm Im }W_a\cr {\rm Im
}W_0 }\end{pmatrix}_{\!\!\!\!\eta}=-|c|\begin{pmatrix}{0\cr 0\cr {\rm Re}X^\L e_\L- {\rm Re}F_\L
m^\L }\end{pmatrix}\ .\label{dRephi-}\ee


\section{The geometry of the type IIA background}
\setcounter{equation}{0}\label{IIA} 

The DW solution of type IIB discussed in the previous section is
expected to have a mirror symmetric solution in type IIA. For RR
fluxes mirror symmetry merely amounts to exchanging the flux of the RR 
three-form $F_3$ defined in (\ref{fluxes}) with the fluxes of the even
forms
$F_2$ and $F_4$ of type IIA \cite{Taylor:1999ii,Louis:2002ny}. However, for the NS-form $H_3$ the
situation is more involved in that mirror symmetry can relate $H_3$-flux
to the torsion of a geometrical compactification
\cite{Vafa:2000wi,Gurrieri:2002wz} or possibly to non-geometrical
quantities \cite{nongeo}. For
electric NS fluxes\footnote{Let us
recall that we suppress the index ``2'' for the NSNS fluxes, that is
we mean $(e_\L,\,m^\L)\equiv(e^2_\L,\,m^{2\L})$.}  $e_\Lambda$
the IIA mirror symmetric solution
corresponds to compactifications on half-flat manifolds 
$\hat{Y}_{\rm hf}$ \cite{HitchinHF,CS,Gurrieri:2002wz}. More
precisely, in  ref.~\cite{Mayer:2004sd}
it was shown that the DW solution 
takes the form of a warped product 
\be \label{X7} M_{(1,2)}\times_w X_7\ ,\ee 
where  the seven dimensional manifold $X_7$ consists a six dimensional
half-flat manifold $\hat{Y}_{\rm hf}$ which is fibered over
$\mathbb{R}$.  Thus the metric takes the form 
\be ds^2_{(7)}=dy^2+ds^2_{(6)}(y)\ ,\ee
where $ds^2_{(6)}$ is the metric of $\hat{Y}_{\rm hf}$ and $y$ is the
coordinate of  $\mathbb{R}$.

Half-flat manifolds are a special sub-class of manifolds with $SU(3)$
structure. They admit a globally defined spinor which is invariant
under $SU(3)$. The existence of this spinor implies the existence of a
two-form $J$ and a complex three-form $\Omega_\eta$.\footnote{$\Omega$
is only defined up to complex rescaling. Therefore a choice of
normalization is involved in the following. By $\Omega_\eta$ we
denote the three-form constructed from  a normalized spinor or
equivalently a three-form which obeys $\Omega_\eta\wedge\bar
\Omega_\eta = \frac{3i}{4} J^3$.}
For half-flat manifolds $J$
and $\Omega_\eta$ satisfy
the additional conditions \cite{HitchinHF,CS}
\be\label{halfflat}
d J^2 = 0 = d {\rm Im} \Omega_\eta \ .
\ee

When  $\hat{Y}_{\rm hf}$ sits inside $X_7$ 
the non-trivial fibration is expressed by the Hitchin flow equations
\cite{HitchinHF,CS} 
\be\label{Hitchinflow}
\frac12\, \partial_y  J^2=- d\,{\rm Re}\Omega_\eta \ , \qquad
\partial_y {\rm Im} \Omega_\eta = d J\ .
\ee
They precisely ensure that $X_7$ has $G_2$ holonomy which corresponds
to the $N=1$ supersymmetry of the IIB  DW solution.

In this section we suggest a generalization of the type IIA geometric
compactification which also captures the mirror of  non-trivial type
IIB magnetic fluxes  $m^{\Lambda}$. 
More precisely we check that compactifications of
the form (\ref{X7}) where  $X_7$ contains a fibered product of a
six-manifold with $SU(3)\times SU(3)$ structure times the real line
are mirror dual to type IIB DW solutions with electric and magnetic flux.
This generalized  mirror symmetry has recently been suggested in
ref. \cite{GMPT,Grana:2005ny,Benmachiche:2006df,GLW} 
and here we confirm that it  
also holds for the case of the DW solution  constructed in the
previous section.

In order to check this proposal let us briefly summarize the results
of refs.\ \cite{Grana:2005ny,GLW}. It was shown that the most general possible
geometrical compactification of type II string theories involves manifolds with 
$SU(3)\times SU(3)$. Such manifolds are defined by the existence of
two locally inequivalent spinors. Each of them is left invariant by an
$SU(3)$ and thus together they define what is called  an $SU(3)\times
SU(3)$ structure \cite{Witt,JW}. Compactifications on such manifolds lead to
an $N=2$ low energy effective action in four space-time
dimensions. The space of scalar fields is most conveniently expressed
in terms of two pure spinors of $SO(6,6)$ denoted by $\Phi_\pm$.
Geometrically $\Phi_+$ is a sum of even forms while $\Phi_-$ is a sum
of odd forms.
If one projects out all possible massive gravitino multiplets both
$\Phi_+$ and $\Phi_-$ 
enjoy an expansion of the form 
\be
\Phi_+=X^\L\o_\L-F_\L\,\o^\L\ ,\qquad
\Phi_-=Z^A_\eta\a_A-W_{\eta A}\,\b^A\ .\label{p2}
\ee 
The $(\o_\L,\,\o^\L)$ form a (non-degenerate) symplectic basis on the
space of even forms while $(\a_A,\,\b^A)$ form a symplectic basis on
the space of odd forms. They are normalized according to:
\be \int_{Y}\o_\L \wedge\o^\S=\d^\S_\L\ ;\qquad  \int_{Y}\a_A \wedge
\b^B=\d^B_A\ .
\ee
In addition $\Phi_\pm$ satisfy a compatibility condition which in
terms of the expansion \eq{p2} reads \cite{Grana:2005ny,GLW}
\be
(X^\L\bar F_\L-\bar X^\L F_\L) = (Z^A\bar W_{A} - \bar Z^A
W_{A})_\eta\ .
\ee

$\Phi_\pm$ are only defined up to arbitrary rescaling and as shown in
\cite{Grana:2005ny} the low energy effective action or more precisely
the K\"ahler potentials depend on the rescaled sections  
 $(Z^A,\,W_A)$ which are related to $(Z^A,\,W_A)_\eta$ 
precisely by the rescaling \eq{Zrescale}. 
In terms of $(X^\L,\, F_\L)$ and $(Z^A,\,W_A)$ the K\"ahler potentials 
are again given by  \eq{KV} and \eq{KH2}, respectively.
Furthermore, it is possible to choose special coordinates 
where $X^0=-i,\,Z^0=1$ holds and in these coordinates
mirror symmetry is realized by imposing
\cite{Benmachiche:2006df,GLW} 
\be\label{d2}
d\a_0=m^\L\o_\L-e_\L\o^\L\ ,\quad d\a_a=d\b^A=0\ ,\quad
d\o_\L=e_\L\b^0\ ,\quad d\o^\L=m^\L\b^0\ .
\ee
One shows that for type IIA compactifications on 
manifolds obeying (\ref{d2})
spectrum and effective action coincide with that obtained by
compactifying type IIB an Calabi-Yau threefolds with electric and
magnetic NS three-form flux turned on \cite{GLW}. For $m^{\L}=0$ one precisely
obtains the half-flat manifolds discussed above.
In this case 
one has $\Phi_+= e^{B+iJ} $ and $\Phi_-= \Omega_\eta$, where $B$ is
the NS two-form.

What is left to study are the $SU(3)\times SU(3)$ generalizations of 
(\ref{halfflat}) and (\ref{Hitchinflow}) and to show that they
correspond to the DW solutions of the previous section.
{}From a mathematical point of view the generalized flow equations
have been derived in ref.~\cite{JW} and (in our notation) they read
\ba
&& d\,{\rm Im}\Phi_-= d\,{\rm Im}\Phi_+=0\ ,\label{gen1}\\
&&\partial_y{\rm Im}\Phi_+=- d\,{\rm Re}\Phi_-\ ,\label{gen2}\\
&&\partial_y{\rm Im}\Phi_-=  d\,{\rm Re}\Phi_+\ .\label{gen3}
\ea
 
Let us now show that these flow equations together with (\ref{d2})
coincide with the DW solution of the previous section.
We start by computing $d\,\Phi_\pm$ and insert
(\ref{d2}) into (\ref{p2}). This yields
\ba
d\,\Phi_+ &=& (X^\L e_\L -F_\L m^\L)\, \beta^0\ , 
\label{dP1}\\
d\,\Phi_- &=&  |c|^{-1} ( m^\L\o_\L-e_\L\o^\L)\ ,
\label{dP2}
\ea
where $|c|$ is defined in \eq{Zrescale}.
{}From the reality of the right hand side of (\ref{dP2}) we
immediately conclude $d\,{\rm Im}\Phi_-=0$.
Furthermore  $d\,{\rm Im}\Phi_+=0$ coincides with the condition
\eq{dImphi+}.

The next step is to compute $\partial_y {\rm Im} \Phi_\pm$.
Using (\ref{p2}) we arrive at
\ba
&&\partial_y{\rm Im}\Phi_+= (\partial_y{\rm Im}X^\L)\, \omega_\L
- (\partial_y{\rm Im}F_\L)\,\omega^\L\ ,\\
&&\partial_y{\rm Im}\Phi_-=  (\partial_y{\rm Im}Z^A_\eta)\, \alpha_A
-(\partial_y{\rm Im}W_{\eta\,A})\,\beta^A\ . \label{gen3p}
\ea
Changing coordinates according to
\be dy= |c|^{-1} dw\ ,\label{cochange}\ee
we see that $\partial_y{\rm Im}\Phi_+=-d\,{\rm Re}\Phi_-$
precisely corresponds to (\ref{dRephi+}) and 
$\partial_y{\rm Im}\Phi_-=d\,{\rm Re}\Phi_+$
corresponds to \eq{dRephi-}. Thus we have achieved our goal and recovered the type IIB flow
equations from the generalized Hitchin flow equations
\eq{gen1}--\eq{gen3} on the type IIA side.

Our next chore is to compare the superpotentials. In \eq{Wdef} we
learned that $W$ is related to
 the matrix $S_{AB}$ defined in \eq{psi}. Precisely this quantity 
 was computed in \cite{Grana:2005ny} in terms of the pure spinors
$\Phi_\pm$ to be
\be
W \sim e^{\frac12(K_V+ K_H) + \varphi} \int_Y d\Phi_+\wedge\Phi_-=
e^{\frac12(K_V+ K_H) + \varphi} (X^\L e_\L-F_\L m^\L)\
,\label{stringW}\ee
where we used \eq{p2} and \eq{d2}.
Again this type IIA quantity precisely coincides with \eq{Wresult} of
type IIB if we also use \eq{Vdef2}.
Thus the Hitchin flow equations can also be viewed as gradient flow
equations of the form \eq{attz}, \eq{attq} with a superpotential given
by \eq{stringW}.

In summary we just showed that the DW solutions of type IIB 
can be expressed as generalized Hitchin flow equations for 
the two pure spinors $\Phi_\pm$ of a manifold with $SU(3)\times SU(3)$
structure as given in (\ref{gen1})--(\ref{gen3}).

Our final task is to discuss the properties of the seven-dimensional
manifold $X_7$.
As the metric on the DW is flat  and the background 
$M_{(1,2)}\times_w X_7$ solves the string equation of motion, we
expect $X_7$ to be Ricci flat. For half-flat manifolds this was indeed
shown in  refs.\ \cite{HitchinHF,CS,Mayer:2004sd}. In order to discuss the
generalization at hand let us introduce 
the seven dimensional exterior derivative by
\be \hat d=d+dy\,\partial_{y}\, ,\ee  
where $d$ acts on $\hat{Y}_6$ and $\partial_{y}$ is the derivative
with respect to the coordinate of $\mathbb{R}$. 
Furthermore, following \cite{Witt,JW} one can define the generalized forms
$\rho$ and $*\r$
on $X_7$ which are given in terms of $\Phi_\pm$ by
\be
\rho=-{\rm Re}\Phi_+\wedge dy-{\rm Im}\Phi_-\ , \qquad
*\r={\rm Re}\Phi_- \wedge dy+ {\rm Im}\Phi_+\ .\label{startre}
\ee 
$*\rho$ is the Hodge dual of $\rho$ 
with respect to the generalized metric.
As noted in \cite{Witt,JW} the equations \eq{gen1}--\eq{gen3} then correspond to
\be d\r=*d\!*\r=0\ ,\ee
and imply that $X_7$ has an integrable
$G_2\times G_2$ structure and is indeed Ricci-flat.

\section{Conclusions and outlook}\label{concl}

In this paper we studied three-dimensional $N=1$ DW solutions of 
four-dimensional  $N=2$ supergravities
which arise as the low energy limit of type IIB string theory
compactified on Calabi-Yau threefolds in the presence of RR and NS three-form
fluxes. An essential ingredient in our analysis was the newly
constructed $N=2$ supergravity
\cite{Dall'Agata:2003yr}--\cite{Sommovigo:2005fk} 
which includes massive antisymmetric
tensors in the spectrum. The use of this supergravity is necessary whenever
magnetic fluxes are turned on as they render antisymmetric tensors in
the type IIB spectrum massive. In this respect we generalized the
previous analysis of refs.\ \cite{Behrndt:2001mx,Mayer:2004sd} and
consistently included magnetic fluxes.
We further showed that the $N=2$ scalar fields vary according to a set of
gradient flow equations and explicitly determined their solution in
terms of the fluxes.

The second aspect of the paper dealt with the type IIA mirror
symmetric DW solutions. Here we used the results of
\cite{Grana:2005ny,Benmachiche:2006df,GLW} and showed that the flow
equations of type IIB have a mirror dual which is purely geometrical
and can be understood as a set of generalized Hitchin flow equations
for a particular class of manifolds with $SU(3)\times SU(3)$ structure
\cite{JW}. As in refs.~\cite{HitchinHF,CS,JW} these flow equations do
have a seven-dimensional interpretation and can be viewed as arising
from fibering a six-dimensional manifold with  $SU(3)\times SU(3)$
over the real line and demanding an integrable $G_2\times G_2$
structure of the resulting
seven-dimensional manifold. 

\vskip 2cm

\subsection*{Acknowledgments}

This work is supported by 
GIF -- The German-Israeli-Foundation under Contract No. I-787-100.14/2003,
DFG -- The German Science Foundation, the
European RTN Programs MRTN-CT-2004-005104,
MRTN-CT-2004-503369 and the
DAAD -- the German Academic Exchange Service. 

We have greatly
benefited from conversations and correspondence with Gabriel Lopes
Cardoso, Mariana Gra\~na, Thomas Grimm,
Peter Mayr, Thomas Mohaupt, Daniel Waldram, Frederick Witt and Marco Zagermann.

\vskip 2cm

\appendix

\noindent
{\bf \Large Appendix}

\section{The scalar $\s$--model of $N=2$ supergravity}\label{th}
\setcounter{equation}{0}

In this appendix we record some further details of the scalar fields
in $N=2$ supergravity. They can be viewed as the coordinates
of some target space geometry which is constrained 
by $N=2$ supersymmetry. In particular the complex scalars of the
vector multiplets lead to a special K\"ahler geometry
while the scalars in the hypermultiplets span a quaternionic manifold
\cite{Andrianopoli:1996cm}. 
Let us discuss both geometries in turn.

\subsection{Special K\"ahler geometry of the vector multiplets}
The complex scalars $t^i,\ i= 1, \ldots, n_V$ belonging to the $n_V$
vector multiplets span a special
K\"ahler geometry. That is their $\sigma$-model metric is a K\"ahler
metric with a K\"ahler potential
\be\label{KV}
K_V= -\ln i\Big[\bar X^\L F_\L - \bar F_\L X^\L\Big]\ ,\quad \
\L = 0, \ldots, n_V\ .
\ee
$X^\L(t)$ and $F_\L(t)$ depend holomorphically on the scalars $t^i$
and are related to the covariantly holomorphic section $V$ introduced in
\eq{Vdef} by
\be\label{hsc}\label{Vdef2}
V = (L^\L, M_\L) = e^{K_V/2} (X^\L, F_\L)\ .
\ee 

For Calabi-Yau compactifications $F_\L=\partial_\L{F}(X)$ is the
derivative of a 
prepotential ${F}$.
In the large volume or large complex structure limit ${F}$ is given by 
\be{F}(X)=-\frac{1}{3!}c_{ijk}\frac{X^iX^jX^k}{X^0}\ , \qquad
i= 1, \ldots, n_V\ ,
\label{ppotvect}\ee
where the $c_{ijk}$ are constants.
A particular set of coordinates, called special coordinates, is given by
\be \label{specialc} t^i\equiv b^i+i v^i=\frac{X^i}{X^0}\ .\ee
In these coordinates the K\"ahler potential (\ref{KV}) is given by
\be\label{KV2}
K_V= -\ln \Big[\ \frac43\, c_{ijk} v^i v^j v^k\Big]\ .
\ee

\subsection{Geometry of  tensor- and hypermultiplets}

The hypermultiplet geometry is described in terms of real scalar fields
$q^{\hat u}$, $\hat u=1,\cdots, 4n_H$, (here $n_H$ is the number of
hypermultiplets) which span a quaternionic manifold. The metric can be
expressed in terms of a covariantly constant vielbein
$\mathcal{U}^{A\a}\equiv\mathcal{U}^{A\a}_{\hat u}dq^{\hat u}$. 
More explicitly one has 
\be h_{\hat{u}\hat{v}}=\mathcal{U}^{A\a}_{\hat
u}\mathcal{U}^{B\b}_{\hat v}\e_{AB}\mathbb{C}_{\a\b}\ ,\qquad A,B =
1,2\ ,\ee
where $\e^{AB}=-\e^{BA}$ and $\mathbb{C}_{\a\b}=-\mathbb{C}_{\b\a}$ are
the $SU(2)$ and $Sp(2n_H,\mathbb{R})$ invariant
metrics respectively. The quaternionic vielbein obeys
\be \nabla\mathcal{U}^{A\a}\equiv d
\mathcal{U}^{A\a}+\hat\o^{A}_{~B}\wedge\mathcal{U}^{B\a}+\hat\D^{\a\b}\wedge\mathcal{U}^{A\b}=0
\ ,\ee
where $\hat\o^{AB}_{\hat u}$, $\hat\D^{\a\b}_{\hat u}$ are the 
$SU(2)$ and $Sp(2n_H,\mathbb{R})$ valued connections.

A set of scalars which parameterizes translational and commuting isometries
can be dualized into a set of $n_T$ antisymmetric rank two
tensors \cite{Dall'Agata:2003yr}. In this case the
remaining scalars $q^u$, $u=1,\cdots, 4n_H-n_T$ will not parameterize a
quaternionic manifold anymore. Instead their $\sigma$-model metric
$g_{uv}$ is given by 
\be
\label{quatm}g_{uv}=h_{uv}-h_{Iu}M^{IJ}h_{Jv}=P_u^{A\a}P_v^{B\b}\e_{AB}\mathbb{C}_{\a\b}\
, \qquad  g^{uv}=h^{uv}\ ,
\ee
where we decomposed the quaternionic metric as
\ba
 h_{\hat{u}\hat{v}} = \begin{pmatrix}{h_{uv}&h_{uJ}\cr
h_{vI}&h_{IJ}}\end{pmatrix}\ ,
\ea
and defined $M^{IJ}$ as  the inverse of $h_{JK}$  
\be M^{IJ}h_{JK}=\d^I_K\ .\ee

The vielbein $P_u^{A\a}$ of the metric $g_{uv}$ defined in \eq{quatm}
can be expressed in terms of the quaternionic vielbein as follows
\be\label{U}P^{A\alpha}_{u}\equiv\mathcal{U}_{u}^{A\alpha}-A^{I}_u\mathcal{U}_{I}^{A\alpha}\
,\qquad P^{u\,A\alpha}\equiv\mathcal{U}^{u\,A\alpha} \ ,\ee
where $A^J_u = h_{Iu}M^{IJ}$. Similarly the connections decompose as
\ba
&&\hat\omega^{AB}_u\equiv\omega^{AB}_u+A^I_u\omega^{AB}_I\ ,\quad\hat\omega^{AB}_I\equiv\omega^{AB}_I;\nn\\
&&\hat\Delta^{\alpha\beta}_u\equiv\Delta^{\alpha\beta}_u+A^I_u\Delta^{\alpha\beta}_I
\ , \quad
\hat\Delta^{\alpha\beta}_I=\Delta^{\alpha\beta}_I\ .
\label{conn}
\ea  
 The new quantities satisfy a certain number of relations
\cite{Dall'Agata:2003yr,Theis:2003jj}  
and here we record only the ones needed in order to derive \eq{condq} and \eq{zerocosm}
\ba
&&\label{quatrel1}(P_u^{A\a}P_v^{B\b}+P_v^{A\a}P_u^{B\b})\mathbb{C}_{\a\b}
=g_{uv}\e^{AB}\ , \\
&&(P_u^{A\a}\mathcal{U}_I^{B\b}+\mathcal{U}_I^{A\a}P_u^{B\b})\mathbb{C}_{\a\b}=0\ ,\\
&&(\mathcal{U}_I^{A\a}\mathcal{U}_J^{B\b}+\mathcal{U}_J^{A\a}\mathcal{U}_I^{B\b})\mathbb{C}_{\a\b}
=M_{IJ}\e^{AB}\ , \\
&&\mathcal{U}_{I\a}^{\ (A}P_u^{B)\a}=\frac 12\nabla_u\o_I^{AB}\ .
\label{quatrel4}
\ea
The covariant derivative $\nabla_u$ is defined with respect to the
reduced  connection $\o_u^{AB}$, $\D_u^{\a\b}$. 

The convention for raising and lowering the symplectic indices is as follows
\ba
\e_{AB}T^B=T_A\ ,&&T_B\e^{BA}=T^A\ ,\\
\mathbb{C}_{\a\b}T^\b=T_\a\ ,&&T_\b\mathbb{C}^{\b\a}=T^\a\ .
\ea
 
\subsection{Quaternionic geometry in Calabi-Yau compactifications}
So far we only discussed the geometry
as it appears in general in $N=2$ supergravity. In Calabi-Yau
compactifications of either type IIA or type IIB string theory only a
special class of quaternionic geometries, termed `dual quaternionic
geometries', arise at the tree level \cite{CGF}.
This is basically a consequence of mirror symmetry and states that 
 the quaternionic manifold of real dimension $4n_H$  necessarily has a special 
K\"ahler submanifold of real dimension $2n_H$ which is spanned by the
geometrical moduli. The remaining $2n_H$ scalar fields then arise from
the RR sector.

Let us be slightly more explicit.
A Calabi-Yau manifold has a geometrical moduli space $\mathcal{M}$ which is product of a
component $\mathcal{M}_{\rm k}$ spanned by the
deformations of the K\"ahler form and a component $\mathcal{M}_{\rm
cs}$ spanned 
by the deformations of the
complex structure
\be \mathcal{M} = \mathcal{M}_{\rm k} \times\mathcal{M}_{\rm cs}\ .\ee
Each component is a special K\"ahler geometry
with a K\"ahler potential of the form 
\eq{KV}, i.e.\ a  K\"ahler potential which can be 
characterized by a holomorphic prepotential.

In compactifications of type IIA the 
deformations of the K\"ahler form reside in vector multiplets while 
the deformations of the
complex structure are members of the hypermultiplets. In type IIB the
situation is exactly reversed and the K\"ahler moduli sit in
hypermultiplets while the complex structure moduli populate the vector
multiplets. In both cases the geometrical moduli in the
hypermultiplets combine with the scalar field from the RR sector
to span the full quaternionic geometry. 

Since we are discussing both type IIA and type IIB compactifications
in the main text we choose to denote the scalar fields in the vector
multiplets by $t^i$ irrespective of their Calabi-Yau origin as 
K\"ahler or complex structure deformations. Similarly, we denote by
$z^a$ the geometrical moduli which reside in the hypermultiplets and
which span the special K\"ahler submanifold inside the quaternionic
manifold. Their K\"ahler potential we denote as
\be\label{KH2}
K_H= -\ln i\Big[\bar Z^A W_A - \bar W_A Z^A\Big]\ ,\quad \
A = 0, \ldots, n_H\ ,
\ee
where  $W_A(Z)$ is the second holomorphic prepotential.
In the large volume or large complex structure limit $K_H$ reduces to
\be\label{KH3}
K_H= -\ln \Big[\ \frac43\, d_{abc} \lambda^a \lambda^b \lambda^c\Big]\ , 
\ee
where 
\be \label{scH}
z^a=\s^a+i\l^a=\frac{Z^a}{Z^0}\ee
are the special coordinates in this
sector.

Finally let us also record the relation with the conventions used in
ref.\ \cite{Ferrara:ik}. In this paper the quantities $\hat K$ and
$\tilde K$ are used which are related to the quantities used in this
paper by
\be e^{-\hat K}=2e^{-K_H}\ , \qquad e^{-\tilde K} =  e^{-2\varphi}\ , \ee
where $\varphi$ is the four-dimensional dilaton.
Finally, the ten-dimensional dilaton  can be expressed as
\be Im\t=4e^\frac{\hat{K}-\wk}{2}\ .\label{newcoord}\ee


\section{Explicit solution of the flow equations}\label{soluz}\setcounter{equation}{0}

In this appendix we derive the explicit solution of the vector multiplets flow equation.

The formal integration of equation \eq{hitch1} is trivial and gives:
\be\begin{pmatrix}{Y^\L-\overline{Y}^\L\cr
\mathcal{F}_\L-\overline\mathcal{F}_\L
}\end{pmatrix}=-i\begin{pmatrix}{m^\L\cr e_\L}\end{pmatrix}x+\begin{pmatrix}{K^\L\cr
K_\L}\end{pmatrix}\ .\label{hitch1int}\ee
Imposing \eq{tadpole}, \eq{dezuz} on \eq{hitch1int} one obtains the condition:
\be K^\L e_\L-K_\L m^\L=0 \label{tadpoleK}\ee
{}From the normalization $Y^0=-\frac i2$ we infer $K^0=1$. 
Explicit integration of \eq{hb}--\eq{h0} yields
\ba
\label{b}b^i\!\!\!&=&\!\!\!m^ix+K^i\\
\label{vv}c_{ijk}v^jv^k\!\!\!&=&\!\!\!c_{ijk}m^jm^k\,x^2+2(c_{ijk}m^jK^k+e_i)\,x+c_{ijk}K^jK^k+2K_i
\ea 
Reinserting \eq{b} and \eq{vv} back into \eq{hi}, \eq{h0} and making use of \eq{tadpoleK} one obtains the following set of constraints on the parameters:
\ba
&&\label{m0}m^0=0\\
&&\label{k0}K^0=1\\
&&\label{mmm} c_{ijk}m^im^jm^k=0\\
&&\label{mmk} c_{ijk}m^im^jK^k+e_im^i=0\\
&&\label{mkk} c_{ijk}m^iK^jK^k+2K_im^i=0\\
&& \frac13c_{ijk}K^iK^jK^k+K_iK^i+K_0=0\\
&& K^ie_i=K^iK_i
\ea
Contacting \eq{vv} with $m^i$ and using \eq{mmm}-\eq{mkk} we further obtain:
\be c_{ijk}m^iv^jv^k=0\label{mvv=0}\ee


\end{document}